\documentclass[epj]{svjour}
\usepackage{graphics}
\usepackage{amsmath}

\begin{document}
\bibliographystyle{unsrt}
\title{Methods for a quantitative evaluation of odd-even staggering effects}
\titlerunning{Methods for a quantitative evaluation of odd-even staggering effects}

\author{Alessandro~Olmi 
\and Silvia~Piantelli
\thanks{\emph{E-mail:} olmi@fi.infn.it, piantelli@fi.infn.it}
}                     
\institute{Sezione INFN di Firenze, Via G. Sansone 1, I-50019 Sesto Fiorentino (FI),
  Italy}
\date{Received: date / Revised version:}
%
\abstract{
Odd-even effects, also known as {\em staggering} effects, are a common feature 
observed in the yield distributions of fragments produced in different 
types of nuclear reactions.
We review old methods, and we propose new ones, for a quantitative estimation 
of these effects as a function of proton or neutron number of the reaction 
products.
All methods are compared on the basis of Monte Carlo simulations.
We find that some are not well suited for the task, the most reliable ones
being those based either on a non-linear fit with a properly oscillating
function or on a third (or fourth) finite difference approach.
In any case, high statistic is of paramount importance to avoid that
spurious structures appear just because of statistical fluctuations in the
data and of strong correlations among the yields of neighboring fragments.
} 

\maketitle

\section{Introduction} \label{sec:intro}

An odd-even effect in the yield distributions of final 
reaction products has been observed since long time in a variety of 
nuclear reactions.
This effect (often called {\em staggering}) usually consists in an enhanced 
production of even nuclear species with respect to the neighboring odd ones:
for example, in the case of charge distributions, even-Z species are 
produced more abundantly than the neighboring odd-Z ones\footnote{
The opposite effect, i.e. enhancement of odd-$Z$ species and reduction
of even ones, is rare and is called {\em antistaggering}.}.

The staggering effect was first investigated long time ago in the fission 
of actinide nuclei, mainly induced by low-energy neutrons (see e.g.
\cite{Runnalls69,Tracy72,Clerc75,Siegert76,Tsekhanovich99,Schmidt01,Tsekhanovich04,Naik07} 
and references therein) and it was attributed to the extra energy required 
to break a pair of protons or neutrons \cite{Rejmund00}.
Experimentally the odd-even effect in fission was found to be less
pronounced for neutrons than for protons.
Many investigations were devoted to finding a systematic parametrization
of the {\em normal\ } shapes, against which experimental results could be 
compared to extract the odd-even effects. 
At very low excitation energies, the yield distributions of fission 
fragments are bell shaped, therefore the 
normal distributions were assumed to be Gaussians \cite{Wahl62}.
The odd-even effect appeared as a superimposed saw-tooth modulation.

Having as a reference these semi-empirical shapes,
the staggering of low-energy fission was usually quantified by a single value 
(called {\em global} staggering) extracted from the whole fragment distribution.
Nowadays staggering effects are studied in a variety of different reactions, 
therefore no a priori shape of the distribution can be assumed
and only {\em local} values of the staggering can be extracted from small 
adjacent regions of the yield distributions.
An extensive review of different quantitative methods used for estimating 
local and global average values of the odd-even effects in fission 
was given by G\"onnenwein \cite{Gonnenwein92}.

For a quantitative analysis,
Amiel and Feldstein \cite{Amiel74,Amiel75} parametrized the odd-even
effects in fragment distributions by assuming that the yields of even-$Z$
fragments are intensified by a factor $(1+\Delta)$ with respect to the
smooth distribution $Y_\mathrm{sm}(Z)$
and those of odd-$Z$ fragments are 
reduced by a factor $(1-\Delta)$, with $\lvert \Delta \rvert < 1$.
The normal behavior is therefore represented by the arithmetic mean 
between odd and even yield values.
The generalization, to take into account odd-even effects as a function 
of both proton and neutron numbers (usually assumed to be independent 
from each other), leads to the expression
\begin{equation}
 Y(Z,N)=  Y_\mathrm{sm}(Z,N) \; (1 \pm \Delta_Z) \; (1 \pm \Delta_N)
                                                          \label{eq:amiel}
\end{equation}
where $ Y_\mathrm{sm}(Z,N)$ is the assumed smooth behavior without staggering; 
the sign `$+$' is used for even Z (or N), the sign `$-$' for odd ones.
$\Delta=0$ means no odd-even effects, while $\Delta\!>0$ indicates enhanced
yields of even species and reduced yields for odd species
(vice versa for $\Delta\!<0$).

Another approach by Wahl \cite{Wahl80} led to a slightly different parametrization 
\begin{equation}
 Y(Z,N)  =  Y_\mathrm{sm}(Z,N) \;  (F_Z)^{\pm 1}  \; (F_N)^{\pm 1} \;,
                                                         \label{eq:wahl}
\end{equation}
\noindent   
with $(F)^{\pm 1}=e^{\pm D}$, which can be cast also in the form
\begin{equation}
    \ln(Y(Z,N)) = \ln(Y_\mathrm{sm}(Z,N)) \; \pm D_Z \; \pm D_N   \label{eq:wahlog}
\end{equation}
where again the sign `$+$' is used for even Z (or N) and the sign `$-$' 
for odd ones.
In this case $F$=1 ($D$=0) corresponds to no odd-even effect, while
$F\!>$1 ($D\!>$0) indicates enhanced yields for even species
and reduced yields for odd species; vice versa for $F\!<$1 ($D\!<$0).
In the parametrization by Wahl, the smooth behavior 
is represented by the geometric mean between odd and even yield values.

Odd-even effects were discovered also in light complex 
fragments produced in high-energy fragmentation and spallation reactions
\cite{Poskanzer71,Webber90,Knott96,Zeitlin97,Ricciardi04,Napolitani07,Napolitani11}.
The staggering consisted again in an intensification of the yields 
of even charges Z with respect to odd ones, although the reaction mechanism,
the shape of the charge distributions and the investigated mass region were 
obviously different from those of low-energy fission experiments.

Recently, staggering effects have been observed also in heavy ion 
collisions from few MeV/nucleon up to Fermi energies
(15 $\leq$E/A$\leq$ 50 MeV/nucleon)
\cite{Yang99,Winchester00,Geraci04,DAgostino11,Ademard11,Lombardo11,DAgostino12,Casini12,Piantelli13}.
These experiments have stirred renewed interest in staggering phenomena.
Indeed, in order to study the symmetry energy 
\cite{Colonna05,Raduta05,Huang10,Su11}, one needs to 
estimate the primary isotopic 
distributions, which can be reliably reconstructed only if the effects of 
secondary decays are small or sufficiently well understood.

The staggering has been usually ascribed to nuclear structure effects
that manifest in the decay of the excited reaction products or already 
in the reaction mechanism, if part of the reaction proceeds through low 
excitation energies \cite{Ademard11}.
However, in collisions at intermediate energies the preferred interpretation 
is that the odd-even staggering effect depends on the structure of the nuclei
produced near the end of the evaporation chain \cite{Ricciardi04,DAgostino11}.
The relationship between nuclear structure and odd-even yield staggering in
nuclear reactions is an active field of research.
Indeed nice results are emerging in well-chosen cases \cite{Mei14}.
However the general case is complex and still not completely clarified,
as it involves not only the nuclear structure of the final nuclei, but
also the level densities and population probabilities of the parent nuclei.

Without loss of generality, we will hereafter refer to the charge
distributions and to the staggering in charge $Z$, but of course exactly 
the same arguments are valid for neutrons too.
Often the presence of odd-even staggering in $Z$ can be visually appreciated 
by simple inspection of the charge distributions, especially in regions
where the effect is prominent. 
For regions with reduced staggering, or in the presence of steep variations 
of the yield as a function of charge, the visualization is more difficult and
some specific handling becomes necessary to highlight the existence of
staggering effects. 

An objective treatment of the data becomes mandatory if one wants to make a
quantitative comparison of different experiments or different nuclear systems. 
Moreover, in contrast with fission fragment studies, for other reaction 
mechanisms the shape of the yield distribution $ Y_\mathrm{sm}$ has to be 
deduced from the data with some procedure.
The first one, introduced by Tracy et al. \cite{Tracy72} in 1972,
has been widely used for long time (see, e.g., \cite{Ricciardi04}), but 
other treatments have been proposed or can be devised.

In this paper we examine some methods that could be used to
put into evidence the presence of the staggering and to quantitatively 
estimate the magnitude of the phenomenon, without relying
on an a priori knowledge of the yield distribution.
These methods are briefly described in sect. \ref{sec:methods} and
the simulation in sect. \ref{sec:simul}.
Section \ref{sec:results} presents the results and a comparison of 
the different methods.

We note that some of the methods presented in this paper
are commonly used for studying odd-even staggering on other nuclear
quantities, the most notable of all being nuclear masses and binding energies
(see e.g. \cite{Jensen84,Hove13} and references therein).
The distinctive feature of fragment yields is that odd-even effects are here
a perturbation superposed on a quantity (the yield) that, depending on the
reaction mechanism, may undergo large and rapid changes in magnitude over
rather restricted regions of $Z$.

\section{Smoothing methods} \label{sec:methods}

When analyzing odd-even effects, the main assumption of eqs.
\eqref{eq:amiel}-\eqref{eq:wahlog} is that the experimental yield $Y(Z)$
can be factorized into the product of the smooth yield $Y_{\mathrm{sm}}(Z)$
(without staggering effects) multiplied by a staggering factor.
In heavy-ion reaction, where $Y_{\mathrm{sm}}(Z)$ is unknown, 
only a local staggering can be determined.
The main, essential assumption is that the staggering varies with 
$Z$ in a sufficiently {\em gradual} way to be considered -- to a good 
approximation -- {\em constant} over an interval of a few $Z$ values.

The smoothing procedures used for estimating  $\Delta_Z$ or $D_Z$ from 
the data are listed below 
(each with the corresponding abbreviation that will be used in the paper) 
and some of them are schematically illustrated in figs. \ref{fig1}
and \ref{fig2}.
For more clarity, the procedures that are applied directly to the experimental
yields will be distinguished from those that are applied to their logarithms
by prefixing ``Y-'' in the first case and ``LY-'' in the second case.
In both figures, the open squares joined by dotted lines indicate the input
cross section without staggering $Y_\mathrm{sm}(Z)$, while the open crosses 
joined by long dashed lines represent what from now on we call the
{\em experimental} cross section $Y(Z)$, because it simulates the outcome of
an experiment
(for illustration purposes no statistical fluctuations are applied 
in these two figures).
The staggering effects are modelled {\em \`a la} Amiel and Feldstein
\cite{Amiel74,Amiel75}, with a constant value of 0.1 for $\Delta_Z$.
The dotted and long dashed lines are just a guide to the eye, joining the points.

\begin{enumerate}
\setlength{\itemsep}{2mm}

\item {\em Linear interpolation (LIN or Y-2DI).}
If the shape of the charge distribution is sufficiently smooth, the simplest 
estimate of $Y_{\mathrm{sm}}(Z)$ is the average between the original 
experimental distribution and the same distribution shifted, 
back and forth, by just one Z unit.
This procedure tends to eliminate any periodical oscillation with period 2$Z$.
Practically, the distribution $\mathcal{Y}(Z)$, 
which is an estimator of the unknown distribution $Y_\mathrm{sm}(Z)$,
can be obtained from the experimental one with a linear interpolation:
\begin{equation}
  \mathcal{Y}(Z) =  \tfrac{1}{4} \big( Y(Z\!-\!1) + 2\; Y(Z) + Y(Z\!+\!1) \big).
                                                             \label{eq:lin}
\end{equation}
In fig. \ref{fig1}(a), $\mathcal{Y}$ is estimated by means of the solid 
line joining the yields $Y_{Z-1/2}$ and $Y_{Z+1/2}$  (open diamonds)
obtained by pairwise averaging the experimental data.
For each $Z$, only the measured yields of three 
consecutive elements (from $Z$-1 to $Z$+1) are needed.
The staggering parameter $\lvert \Delta_Z \rvert$ is given by
$\lvert Y(Z)-\mathcal{Y}(Z) \rvert \,/\,\mathcal{Y}(Z)$
and its effect is indicated by the arrows pointing from the experimental
$Y(Z)$ (crosses) to $\mathcal{Y}(Z)$ (circles).
This method is similar to that applied by Zeitlin \cite{Zeitlin97} 
(and later by others \cite{Iancu05,Cheng12}) in high-energy 
fragmentation studies.
Note that in his original version, Zeitlin \cite{Zeitlin97} took as a reference 
for each given $Z$ the line (fine dotted in figure) joining $Y(Z-1)$ and $Y(Z+1)$:
in this way he obtained an odd-even effect (distance from cross to triangle, 
shown only for $Z$=7) exactly twice as large as the present one 
(distance from cross to full circle).

\item {\em Fit-Interpolation over five points (Y-2DIF).}
In this paper we propose a possible improvement of the previous method
by taking into account also the yields of the two neighboring points
$Y(Z-2)$ and $Y(Z+2)$.
These five yields are pairwise averaged to produce four intermediate
values $ Y_{Z-3/2}, Y_{Z-1/2}, Y_{Z+1/2}$ and $Y_{Z+3/2}$ (open diamonds), 
which are used to perform a parabolic fit (short dashed line).
The value of the fit in the central point $Z$ gives the estimated 
$\mathcal{Y}(Z)$ and the difference with respect to the measured value 
is used to estimate $\Delta_Z$.

\begin{figure}
 \resizebox{0.450\textwidth}{!}{%
  \includegraphics{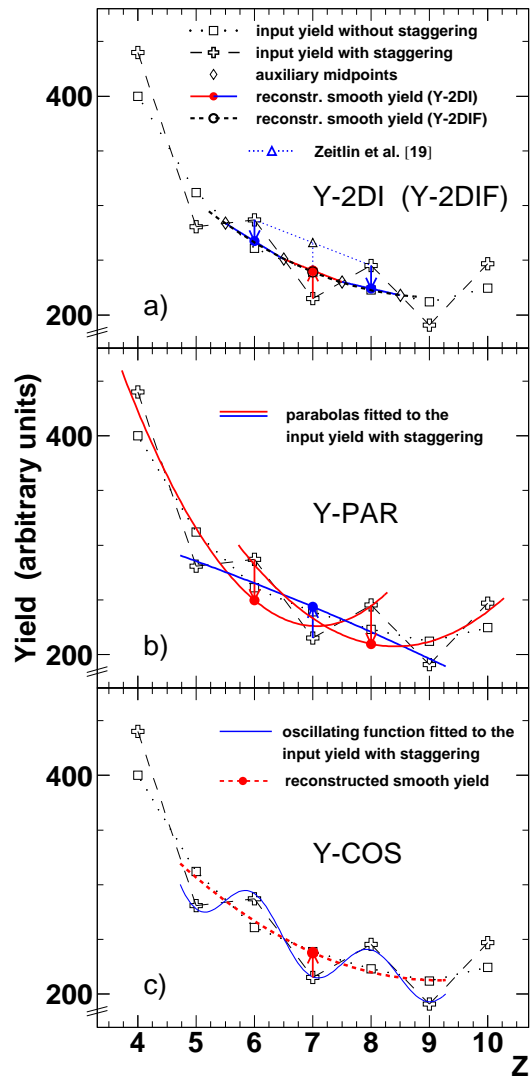}}
  \caption{(color online) Methods Y-2DI and Y-2DIF (a), Y-PAR (b) and Y-COS (c).
   The smooth input distribution (squares) is multiplied (crosses) by the
   staggering factor $(1+ (-1)^Z \Delta_Z)$.
   No statistical fluctuations are applied.
   The full lines represent the constructions used to estimate the 
   smoothed yields $\mathcal{Y}(Z)$ (full circles), the arrows show the
   magnitude of the estimated staggering (see text).
   Panel (a) shows also the method of Zeitlin \cite{Zeitlin97} 
   (triangle and fine dotted line).
    }
\label{fig1}       
\end{figure}

\item {\em Parabolic fit (Y-PAR).}
A new smoothing method was recently suggested in Ref. \cite{DAgostino11}, 
based on a fitting procedure too.
For each measured value of the yield $Y(Z)$, the smoothed value 
$\mathcal{Y}$(Z) is estimated by fitting a para\-bolic function,
\begin{equation}
 \mathcal{Y}(Z)  = a\; Z^2 + b\; Z + c,                \label{eq:par}
\end{equation}
directly to the measured yields over five consecutive values of $Z$, 
from $Z-2$ to $Z+2$.
Figure \ref{fig1}(b) illustrates this method showing three curves,
two for the interpolation around even $Z$ (upward parabolas, red online) and 
the other one around an odd $Z$ (downward parabola, blue online).
The effect of $\Delta_Z$ is again indicated by the arrows pointing from
the experimental $Y(Z)$ (crosses) to $\mathcal{Y}(Z)$ (full circles).
The three parabolas show the peculiarity of this method.
Due to the presence of the staggering, the experimental points have an
up-and-down behavior that is obviously not well fitted by a parabola.
Therefore the estimated effect of staggering (arrows in figure) is enhanced.
In fact, with respect to the true yields without staggering
$Y_{\mathrm{sm}}(Z)$ (open squares), the estimated values of $\mathcal{Y}(Z)$ 
(full circles) lie systematically below for even $Z$ values, 
and above for odd ones.
This amplifies the staggering effect, but does not make the 
method more sensitive, because also the statistical errors are enhanced by 
the same factor. At the same time one has to be aware that this  
amplification hinders a proper quantitative comparison with the results of 
other experiments, evaluated with other methods.

\item {\em Parabolic fit with oscillations (Y-COS).}
In this paper we propose an improvement of the previous method
by taking into account the fact that actually, due to staggering, 
the data oscillate around the assumed smooth behavior.
The improvement is accomplished by using a fit function that consists of 
a parabola multiplied by any properly oscillating function with 
period 2$Z$ and amplitude $d_o$, like for example 
\begin{eqnarray}
 Y_\mathrm{fit}(Z)
 & = &  \big(a_o\; Z^2 + b_o\; Z + c_o\big)
        \; \big(1 + d_o\; \cos(\pi Z) \big) \nonumber \\
 & = &  \makebox[8mm][c]{} \mathcal{Y}(Z)~~~~\big(1 + d_o\; \cos(\pi Z)\big). 
      \label{eq:cos}
\end{eqnarray}
Of course any other oscillating function (like a rectangular or a triangular one)
with period 2$Z$ and proper phase, would have done an equally good job; we choose
$\cos(\pi Z)$ because it is simple and easy to implement in an analysis code.
At the cost of a non-linear fitting procedure\,\footnote{
  All non-linear fits were performed by means of \textsc{Minuit} \cite{minuit}
  (with successive calls to the routines \textsc{Migrad}, \textsc{Hesse} 
  and \textsc{Minos}) and always reached convergence.}
and by adding a fourth parameter, one obtains a better and more sensible 
fit (in fact, $Y_\mathrm{fit}(Z)$ {\em does} now take into account the 
oscillation of the data).
There is also an additional bonus: the $\chi^2$ acquires statistical 
significance and indeed its distribution agrees very well with that 
expected for the $\chi^2$ function with one degree of freedom.
Two variants are possible: 
(a) deriving $\Delta_Z$ from the difference between the estimated
   $\mathcal{Y}(Z)$ and the measured $Y(Z)$, or
(b) directly using the fourth parameter of the fit $\Delta_Z \equiv d_o$.
We will call these two variants Y-COSa and Y-COSb, respectively.
The result of just one fit in the five-point interval around the charge $Z$=7
is shown by the thin solid line in fig. \ref{fig1}(c).
The thick dashed line shows the parabolic part $\mathcal{Y}(Z)$ in eq. \eqref{eq:cos}.
The estimate of $\Delta_Z$ obtained with variant (a) is indicated once more by
the arrow pointing from the experimental $Y(Z)$ (open cross)
to $\mathcal{Y}(Z)$ (full circle).
It is apparent that now the circle represents a much better estimate of 
$Y_\mathrm{sm}(Z)$ than in the previous method.

\begin{figure}[t]
 \resizebox{0.45\textwidth}{!}{%
  \includegraphics{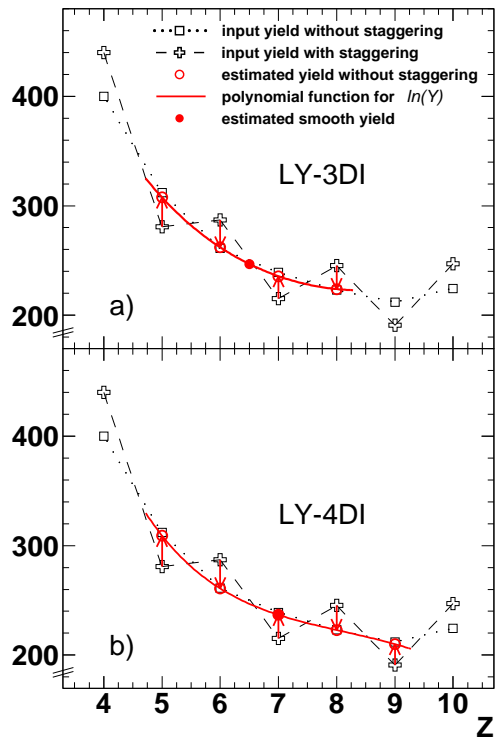}}
  \caption{(color online) Illustration of the methods
   (a) LY-3DI and (b) LY-4DI. 
   Open squares and crosses are the same as in fig. \ref{fig1}. 
   The effect of the estimated staggering is shown
   by the arrows pointing from the experimental data (crosses) to the 
   reconstructed smooth yields (open circles).
   The full lines are the assumed curves that best fit the open circles
   and the full circle gives the so estimated value of $Y_\mathrm{sm}$:
   see text for details.
   Note the expanded vertical scale with suppressed origin.
    }
\label{fig2}       
\end{figure}

\item {\em Log-Third difference method (LY-3DI).}
This is actually the oldest method. First proposed by Tracy \cite{Tracy72}, 
it has been widely used in the past and also in 
recent publications (see e.g. \cite{Yang99,Ricciardi04}).
It relies on finite difference calculus (see Appendix A)
and in particular it uses the third difference of the natural logarithm of
the measured yield $L(Z)= \ln(Y(Z))$ over four consecutive $Z$ values:
\begin{eqnarray}
    D^{(3)}_{Z+1/2} & =  (-1)^{Z}\; \tfrac{1}{8}\; & \{L(Z\!+\!2)-3\; L(Z\!+\!1) 
                                                                \nonumber \\
   &  &  \; +3\;L(Z)- \;L(Z\!-\!1)   \}. \makebox[5mm][c]{}     \label{eq:l3di}
\end{eqnarray}
The basic assumption is that, without staggering effects, the shape of 
the logarithm of the yield, $L(Z)$, over the considered interval of 
four $Z$ values could be described by a parabola
\begin{equation}
            L(Z) = a_{\,T}\; Z^2 + b_{\,T}\; Z + c_{\,T} .       \label{eq:Ltracy}
\end{equation}
The justification by Tracy was that the smooth behavior of $Y_\mathrm{sm}(Z)$ 
over a small interval of $Z$ can be well described by a piece of a Gaussian.
It is worth noting that this does {\em not} necessarily imply that the 
whole charge distribution needs to be Gaussian-shaped, as it was usually
assumed in old-time low-energy fission studies\,\footnote{
  Actually, as either sign is allowed for the coefficient $a_{\,T}$
  in eq. \eqref{eq:Ltracy}, the shape of $Y(Z)$ is not always a piece
  of a Gaussian.}.
One can easily verify that with the functional form of eq. \eqref{eq:Ltracy}
(and with no staggering effects) the following identity holds
\begin{equation}
    ^{\mathrm{ns}}D^{(3)}_{Z+1/2} \equiv 0.            \label{eq:D3_0}
\end{equation}
If now a fixed quantity $D_Z$ is alternatively added to $L(Z)$ for even $Z$
and subtracted for odd ones, then one obtains $D^{(3)}_{Z+1/2 } \equiv D_Z$.
One can easily recognize Wahl's parametrization \cite{Wahl80} of the
staggering with $D_Z\!\equiv \ln(F_Z)$, see eq. \eqref{eq:wahlog}.
The factor $(-1)^Z$ in eq. \eqref{eq:l3di} takes care of the fact that for
staggering the sign of $D_Z$ is usually positive for even $Z$ and negative
for odd ones.
We note also that the result is usually associated with the center of 
the interval and therefore, being $D^{(3)}$ an {\em odd}
finite difference, that center corresponds to a {\em half-integer} $Z$ value. 
An illustration of this method is shown in fig. \ref{fig2}(a),
where the arrows pointing from four consecutive experimental values of 
$Y(Z)$ (crosses) to the corresponding open circles represent the estimated 
staggering effect over that interval and the
solid line is the quadratic polynomial, in the logarithm of the 
yields $L(Z)= \ln(Y(Z))$, that best fits the four open circles.
The central full circle (red online) gives the so estimated
value of $Y_\mathrm{sm}$ at the center of the considered interval.
\newcounter{salvam}
\setcounter{salvam}{\value{enumi}}
\end{enumerate}
\noindent
   The method LY-3DI by Tracy, applied to four consecutive points,
   is just a particular application of the mathematical formalism of
   Finite Difference Calculus, as briefly illustrated in Appendix A.
   This formalism has been widely used and discussed
   in the literature in connection with nuclear structure features,
   like pairing effects in nuclear masses and binding energies
   (see e.g. \cite{Jensen84,Hove13}), but not so much in connection
   with the fragment production yields.
   
   Actually, already the calculation for the first method, LIN,
   produces a 3-point equation that corresponds to the second-order
   difference in the fragment yields.
   For this reason we will henceforth address that method as Y-2DI:
   it represents the simplest useful implementation of this formalism
   (in fact, a first-difference 2-point formula would probably be too rough).

   Other implementations are possible, but -- to our knowledge --
   they were not considered for studying odd-even effects,
   at least not for what concerns the yields of reaction products.
   Applying eq. \eqref{eq:pol_amiel_3} of Appendix A to the yields $Y(Z)$
   and eq. \eqref{eq:pol_wahl_3} to their {\em logarithms} $L(Z)$,
   one obtains additional methods that are briefly sketched
   hereafter.

\begin{enumerate}
\setlength{\itemsep}{2mm}
\setcounter{enumi}{\value{salvam}}

\item {\em Log-Second difference method (LY-2DI).}
For sake of completeness, we mention that a linear method, similar to Y-2DI,
could be applied also to the logarithm of the yields,
leading to the second-difference expression
\begin{equation}
    D^{(2)}_Z  = 
    (-1)^{Z}\; \tfrac{1}{4}\; \{-L(Z\!-\!1)+2L(Z)-L(Z\!+\!1)\}   \makebox[7mm][c]{}
                                    \label{eq:l2di}
\end{equation}
\noindent
which is again an estimator of $D_Z$,
with the superscript indicating the order of the finite difference.
As in the case of Y-2DI, also this method uses three points.

\item {\em Log-Fourth difference method (LY-4DI).}
Much in line with the method of Tracy \cite{Tracy72}, 
in the present paper we consider of great interest also
the fourth difference of $L(Z)$ over five consecutive points:
\begin{eqnarray}
  D^{(4)}_Z& =& \; (-1)^{Z} \; \tfrac{1}{16} \;
       \{L(Z\!-\!2)\;-4\;L(Z\!-\!1)
                                                              \nonumber \\
  &+ & \;6\;L(Z)\;-4\;L(Z\!+\!1)\;+L(Z\!+\!2)\}. 
                                    \makebox[7mm][c]{}   \label{eq:l4di}
\end{eqnarray}

Again the assumption is that, without staggering, $L(Z)$ is
well described by eq. \eqref{eq:Ltracy} and also in this case one 
obtains $^\mathrm{ns}D^{(4)}_Z \equiv 0$.
However, if a quantity $D_Z$ is alternatively added to and subtracted
from $L(Z)$, one finds again $D_Z^{(4)} \equiv D_Z$.
With respect to the method if Tracy, this one uses five points 
(like all previous methods except Y-2DI and LY-3DI) 
and the staggering parameter $D_Z$ can be more physically attributed to 
the {\em integer} $Z$ value at the center of the considered interval.
An additional advantage of LY-4DI over LY-3DI is that eq. \eqref{eq:l4di}
remains valid even if $L(Z)$ has some cubic component (see Appendix A):
\begin{equation}
            L(Z) = a\; Z^3 + b\; Z^2 + c\;Z + d .       \label{eq:Lcubic}
\end{equation}
An illustration of LY-4DI is shown in fig. \ref{fig2}(b), with the
same meaning of symbols and lines as in fig. \ref{fig2}(a).

\item {\em Third difference method (Y-3DI).}
The difference of this method with respect to LY-3DI is that it uses the
fragment yields (instead of their logarithms) of four fragments,
giving a value of $\Delta_Z$ that is again attributed to the half-integer
intermediate value $Z+\tfrac{1}{2}$.
From eq. (\ref{eq:pol_amiel_3}) the staggering parameter $\Delta_z$
is obtained as
\begin{eqnarray}
 \Delta_{Z+1/2}^{(3)} = (-1)^{Z}\; \cdot \makebox[52mm][c]{} \nonumber \\
  \{Y(Z\!+\!2)-3\; Y(Z\!+\!1)+3\;Y(Z)- \;Y(Z\!-\!1)\} \big/   \makebox[6mm][c]{}
      \nonumber  \\
   \{Y(Z\!+\!2)+3\; Y(Z\!+\!1)+3\;Y(Z)+ \;Y(Z\!-\!1)\}   \makebox[7.5mm][c]{} 
    \label{eq:3di}
\end{eqnarray}

\item {\em Fourth difference method (Y-4DI).}
The next implementation with five fragment yields, again based on the
fourth finite difference of eq. (\ref{eq:pol_amiel_3}), reads:  
\begin{eqnarray}
  \Delta_Z^{(4)} & = & (-1)^{Z} \;  \{Y(Z\!-\!2)-4\;Y(Z\!-\!1) \nonumber \\
 & & \; +6\;Y(Z)-4\;Y(Z\!+\!1)+Y(Z\!+\!2)\} \big/  \makebox[5mm][c]{}  \nonumber  \\
 & &        \makebox[9.5mm][c]{}    \{Y(Z\!-\!2)+4\;Y(Z\!-\!1)  \nonumber  \\
 & & \; +6\;Y(Z)+4\;Y(Z\!+\!1)+Y(Z\!+\!2)\}  \makebox[6.5mm][c]{}  \label{eq:4di}
\end{eqnarray}  
\noindent
This method has been extensively employed in nuclear structure
studies of odd-even mass staggering \cite{Hove13}.

\end{enumerate}

Some comments are due, before beginning to examine the merits and
drawbacks of the various methods that can be used to extract the
staggering parameter $\Delta_Z$ (or $D_Z$).
First of all, it is not advisable to use still higher-order finite
differences because they perform weighted averages over an increasing
range of experimental points.
One has to find a compromise between using more points to be less sensitive
to higher derivatives of the smooth yield function, and using less points
to be more sensitive to local variations of the staggering phenomena.
We decided not to go beyond five points, which is the number
commonly used also by the other methods based on fitting procedures.

We also point out that the methods based on fits and those using finite
differences of the fragment yields adopt the staggering parametrization by
Amiel-Feldstein \cite{Amiel74,Amiel75}, while those using finite differences
of the yield logarithms find more natural to adopt that by Wahl \cite{Wahl80}.
We will see that this has only minor consequences.

A further dissimilarity is that the methods relying on fits do first try to
extract, from the data themselves, the smooth yield $\mathcal{Y}(Z)$
(full circles) that best estimates the unknown $Y_{\mathrm{sm}}(Z)$.
This in turn allows one to determine $\Delta_Z$ from the comparison with 
the experimental yield:
\begin{equation}
  \Delta_Z = (-1)^Z \; (Y(Z) - \mathcal{Y}(Z)) \; /\;  \mathcal{Y}(Z)
                                                 \label{eq:Delta}
\end{equation}
(actually also Y-COSb gives $\Delta_Z$ directly from the fit). 
In eq. \eqref{eq:Delta} the factor $(-1)^Z$ takes care again of the
opposite signs for even and odd $Z$ values.

On the contrary, in the methods using the finite difference formalism there is
generally no need to compute $\mathcal{Y}(Z)$, because they give directly
$\Delta_Z$ (or $D_Z$), which is the quantity of physical interest.
In figs. \ref{fig2}(a) and (b) the underlying $Y_\mathrm{sm}(Z)$ 
distributions have been reconstructed just for illustrative purposes:
by subtracting the staggering effect from the experimental yields $Y(Z)$
(crosses), one can deduce the smooth yields (open circles) and draw the curves
(solid lines) corresponding to eqs. \eqref{eq:Ltracy} and \eqref{eq:Lcubic}.

The two parametrizations by Amiel-Feldstein and Wahl are approximately 
equivalent in case of small staggering effects ($\Delta_Z \ll 1$),
being to first order $D_Z \approx \Delta_Z$.
In fact, taking the ratio between the yields for even and odd species
in both parametrizations, one obtains the relationship \cite{Gonnenwein92}
\begin{equation}
 (1+\Delta_Z)/(1-\Delta_Z)=(F_Z)^{\,2} 
    \; \equiv (e^{D_Z})^2           \label{eq:d-f}
\end{equation}
and expanding
$1+2\,\Delta_Z+\mathcal{O}\,(\Delta_Z^2) = 1+2\,D_Z+\mathcal{O}\,(D_Z^2)$.

In case of larger staggering, a fair quantitative comparison of the results
obtained with the two parametrizations may require corrections of higher
order \cite{Gonnenwein92}:
\begin{eqnarray} 
D_Z & = & \Delta_Z + \tfrac{1}{3}\; \Delta_Z^{~~3} 
       +\; \tfrac{1}{5}\; \Delta_Z^{~~5}
       + \mathcal{O}\,(\Delta_Z^{~~7})                  \label{eq:delta_Z} \\
\Delta_Z & = & D_Z - \tfrac{1}{3}\; D_Z^{~~3}
       + \tfrac{2}{15}\; D_Z^{~~5} + \mathcal{O}\,(D_Z^{~~7})   \label{eq:d_Z}
\end{eqnarray}

\section{Simulation} \label{sec:simul}

A comparison of the different methods, to find out their possible 
advantages and disadvantages, is best performed by means of simulations.
In this way one can play with the smooth yield 
distribution $Y_\mathrm{sm}(Z)$ and with the parameter $\Delta_Z$, 
in order to produce a final distribution $Y(Z)$ (including statistical 
fluctuations) that simulates the experimental one.
This allows one to test in how far the different methods are able to
retrieve the original $Y_\mathrm{sm}(Z)$ and 
the genuine odd-even effect $\Delta_Z$.
In the simulations for the present paper, the Amiel and Feldstein
parametrization ($\Delta_Z$) has been adopted and the results of the
LY-2DI, LY-3DI and LY-4DI methods (that deliver $D_Z$) have been
transformed by means of eq. \eqref{eq:d_Z}.

The simulation proceeds in the following steps:
\begin{enumerate}
\item  \label{item:1}
A smooth analytic function $f(x)$ is assumed and its evaluation
for integer $x$ ($x \in N$) gives the {\em discretized } smooth charge 
distribution $Y_\mathrm{sm}(Z)$ without staggering.
\item  \label{item:2}
Following the parametrization by Amiel and Feldstein,  
the charge distribution with staggering is obtained as
$Y(Z) = Y_\mathrm{sm}(Z)\; (1 + (-1)^Z \Delta_z)$, where in principle
$\Delta_z$ could even be a slow-varying function of $Z$.
\item  \label{item:3}
A proper normalization converts the yield $Y(Z)$ into the number of counts
$N_c(Z)$ and then fluctuations are added to $N_c(Z)$, thus producing an
experimental charge distribution that simulates the outcome of an experiment.
Statistical fluctuations are produced by means of the random number generator 
\textsc{Ranlux} \cite{ranlux}.
In principle they should be Poisson-distributed,
but the normalization is chosen to give values of
$N_c(Z)$ large enough, so that the fluctuations can be extracted from a
Gaussian distribution with $\sigma^2(Z) = N_c(Z)$.
\item  \label{item:4}
All the described smoothing methods are applied to the experimental
distribution to obtain values of $\Delta_Z$ as a function of $Z$.
\newcounter{salvai}
\setcounter{salvai}{\value{enumi}}
\end{enumerate}
\noindent
Steps \ref{item:3} and  \ref{item:4} are repeated for a preset number
of times, which in our case is always $10^4$.
\begin{enumerate}
\setcounter{enumi}{\value{salvai}}
\item  \label{item:5}
The propagation of the fluctuations on the individual yields, including 
all possible correlations and non-linearities of the methods, produces 
-- for each $Z$ -- a distribution of $\Delta_Z$ that can 
be well fitted with a Gaussian.
The centroids $\langle \Delta_Z (Z) \rangle$ give the distribution of the 
average result (over $10^4$ distributions) for each method.
\item  \label{item:6}
The results obtained with the different methods are finally plotted and 
examined in detail.
\end{enumerate}
With reference to the above step \ref{item:5}, we emphasize that our 
procedure is equivalent (as we have easily verified with the simulation)
to adding up all $10^4$ replicas and applying the various methods to the 
summed distribution (which has a huge statistic).
The advantage of our procedure is that the widths of the Gaussians, 
$\sigma_{\Delta}$, give the uncertainty on $\Delta_Z$ 
-- at the $1 \sigma$ level -- expected in the analysis of {\em one single} 
experimental distribution.
This is exactly the meaning that will be given to the dashed lines 
in the panels (b) to (h) of fig. \ref{fig:zero} and to the error bars in 
the panels (b) to (m) of figs. \ref{fig:maxmin} to \ref{fig:spline_low}.

Our final aim is to compare the behavior of
the different methods of sect. \ref{sec:methods}, in relation to both
magnitude of the staggering effect and shape of the charge 
distribution, as well as to assess the robustness of the result with 
respect to the statistics of the collected data.

For what concerns the last point, to avoid that intrinsic differences among 
the various analysis methods are obscured by fluctuations, the 
{\em relative} errors on the yields should be kept appropriately small.
In the simulation, where the fluctuations are just of statistical nature,
it is enough to use a reasonably large number of counts $N_c(Z)$~\footnote{
  In a real experiment this may not suffice, as one has to include other
  possible uncertainties, if any, that may affect neighboring charges $Z$ 
  in an independent way (e.g. contaminations, identification uncertainties,
  and so on).}.

One should be aware that the fluctuations on one $Z$ affect 
the estimated value of the staggering parameter $\Delta_Z$ also for 
neighboring charges, and this happens in quite a complicated and 
correlated way that is difficult to estimate without a simulation.
In fact, while in a real experiment one can analyze just one single 
realization of the charge distribution, a big advantage of the simulation 
is that one can generate a large number of replicas of the same charge 
distribution (differing from each other only in the random fluctuations) 
and all correlations come naturally out from the simulation itself.

Finally, the simulation allows one to choose freely the shape of 
$Y_\mathrm{sm}(Z)$ in order to see the response of the different methods 
to the shape of the charge distribution, with special attention to its
regions of non-linearity.

\section{Results} \label{sec:results}

\begin{figure}
\resizebox{0.50\textwidth}{!}{%
  \includegraphics{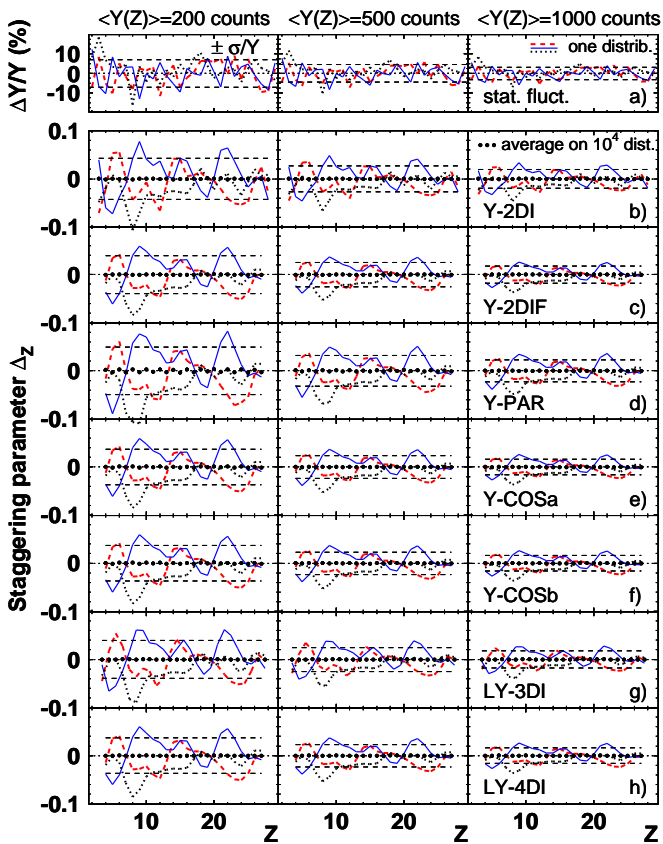}}
  \caption{(color online) Effect of statistical fluctuations for three constant
   charge distributions with 200, 500 and 1000 counts per $Z$ unit
   (first, second and third column, respectively), 
   without staggering ($\Delta_Z = 0$).
   Panels in first row (a): relative statistical  
   fluctuations, $(Y\,$-$\,\langle Y \rangle)/\langle Y \rangle$,
   generated by three different sequences of random numbers
   (full, red dashed and dotted lines), that applied 
   to the constant distributions produce three experimental realizations.
   Panels in rows from (b) to (h): average values (full dots) of the  
   parameter $\Delta_Z$ evaluated over $10^4$ distributions with some of the
   methods illustrated in this paper, compared with $\Delta_Z$ obtained for each 
   of the three realizations (see text).
}
\label{fig:zero}       
\end{figure}

The first thing that we checked is that indeed all methods are unbiased, 
i.e. in absence of staggering they give $\Delta_Z = 0$ within errors.
This is shown in fig. \ref{fig:zero}, where we consider constant 
charge distributions, without staggering, with a statistic of 200, 500 and
1000 counts per $Z$ unit (first, second and third columns, respectively).
In the first row, fig. \ref{fig:zero}(a), the continuous, dashed
(red online) and dotted lines show the relative errors $\Delta Y / Y$
obtained by applying the same three sequences of random errors (step 3 of 
sect. \ref{sec:simul}) to the above mentioned constant distributions:
in fact the structure is the same, the relative errors being
reduced just because of the increasing number of counts.
The horizontal dashed lines 
show the $\pm 1 \sigma$ error band expected for just one realization and
correspond to relative errors $\Delta Y / Y$ of about $\pm\,$7\% 
for $N_c$=200, down to about $\pm\,$3\% for $N_c$=1000.
This means that in one single realization of the charge distribution 
(i.e. in one measurement), for each measured $Z$ there is still a probability of 
$\approx$ 32\% that its measured yield lies outside this band.

\begin{figure*}
\resizebox{0.80\textwidth}{!}{%
  \includegraphics{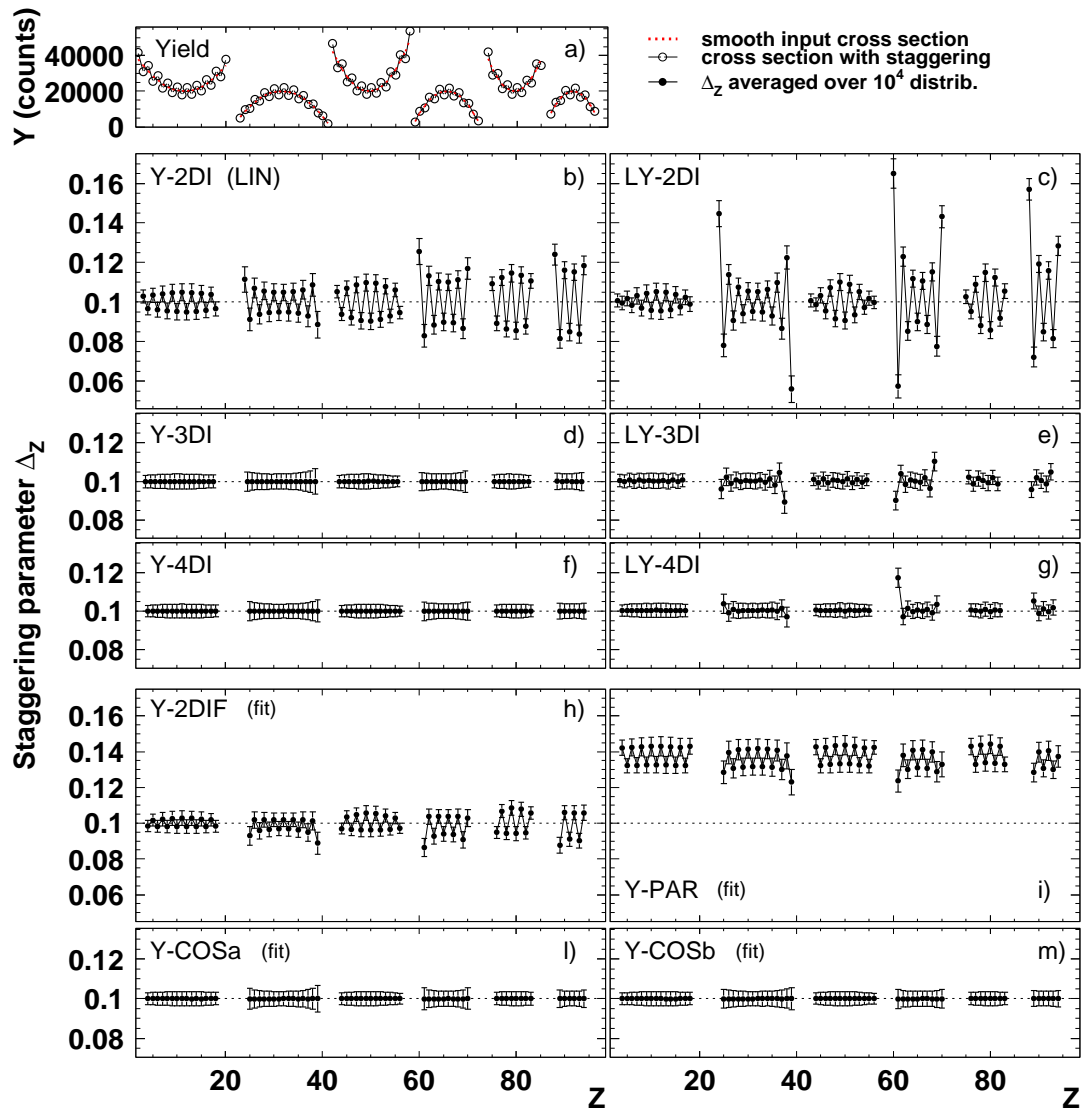}}
  \caption{(color online) 
   Upper panel (a): charge distributions shaped 
   as vertical parabolas (pairwise with the same curvature). 
   Open dots show a constant staggering $\Delta_Z$=0.1 applied to the smooth 
   dashed curves.
   Panels (b)-(m) present the average  $\langle \Delta_Z \rangle$ obtained with
   the methods Y-2DI, LY-2DI, Y-3DI, LY-3DI, Y-4DI, LY-4DI, Y-2DIF, Y-PAR,
   Y-COSa and Y-COSb, respectively, compared with the nominal input value
   (dotted lines).
   The error bars are the $\pm 1 \sigma$ widths of the $\Delta_Z$ distributions.
}
\label{fig:maxmin}       
\end{figure*}

These relative errors may seem rather small, but in the evaluation
of $\Delta_Z$ they are amplified by all examined methods.
In fact propagating these errors from the yields to $\Delta_Z$ leads 
to the $\pm 1 \sigma$ band indicated by the horizontal dashed lines in rows 
(b) to (h) of fig. \ref{fig:zero},
its width amounting for all methods to about $\pm$ 0.04  
when $N_c$=200 and $\pm$ 0.02 when $N_c$=1000~\footnote{
  Similar absolute uncertainties occur also in presence of staggering 
  and cannot be overlooked: for $\Delta_Z$ = 0.1 they correspond to an
  uncertainty of $\approx \pm\,$40\% (20\%) for $N_c$=200 (1000).}.

But this is not the worst part of the story.
The impact of these fluctuations on the estimation of $\Delta_Z$ is 
presented in the lower panels of fig. \ref{fig:zero}, rows from (b) to (h).
One clearly sees that taking the average (step 5 of sect. \ref{sec:simul}) 
over $10^4$ generated replicas, differing only in the fluctuations,
gives $\Delta_Z = 0$ with all methods (full dots).
However, the correlated propagation of the fluctuations of neighboring 
charges gives rise to rather large and nonphysical structures 
(full, dashed and dotted lines),
with different and often opposite phases in different realizations:
one realization may have a bump where another one has a valley.
It is also worth noting that the spurious bumps (or valleys) depend only on the
particular set of random fluctuation and appear to be strictly 
{\em in phase} for all analysis methods that can be applied.

The main conclusions that can be drawn from fig. \ref{fig:zero} are that
(a) fluctuations in the charge distribution are amplified by all methods 
and the values of $\Delta_Z$ for neighboring $Z$ become strongly correlated; 
(b) good statistic (and reduction of other sources of fluctuations) is 
therefore mandatory for any experiment addressing this topic;
(c) structures or trends in $\Delta_Z(Z)$ should be interpreted 
with great caution, after a careful and realistic estimate of their
statistical significance.

\begin{figure*}
\resizebox{0.80\textwidth}{!}{%
  \includegraphics{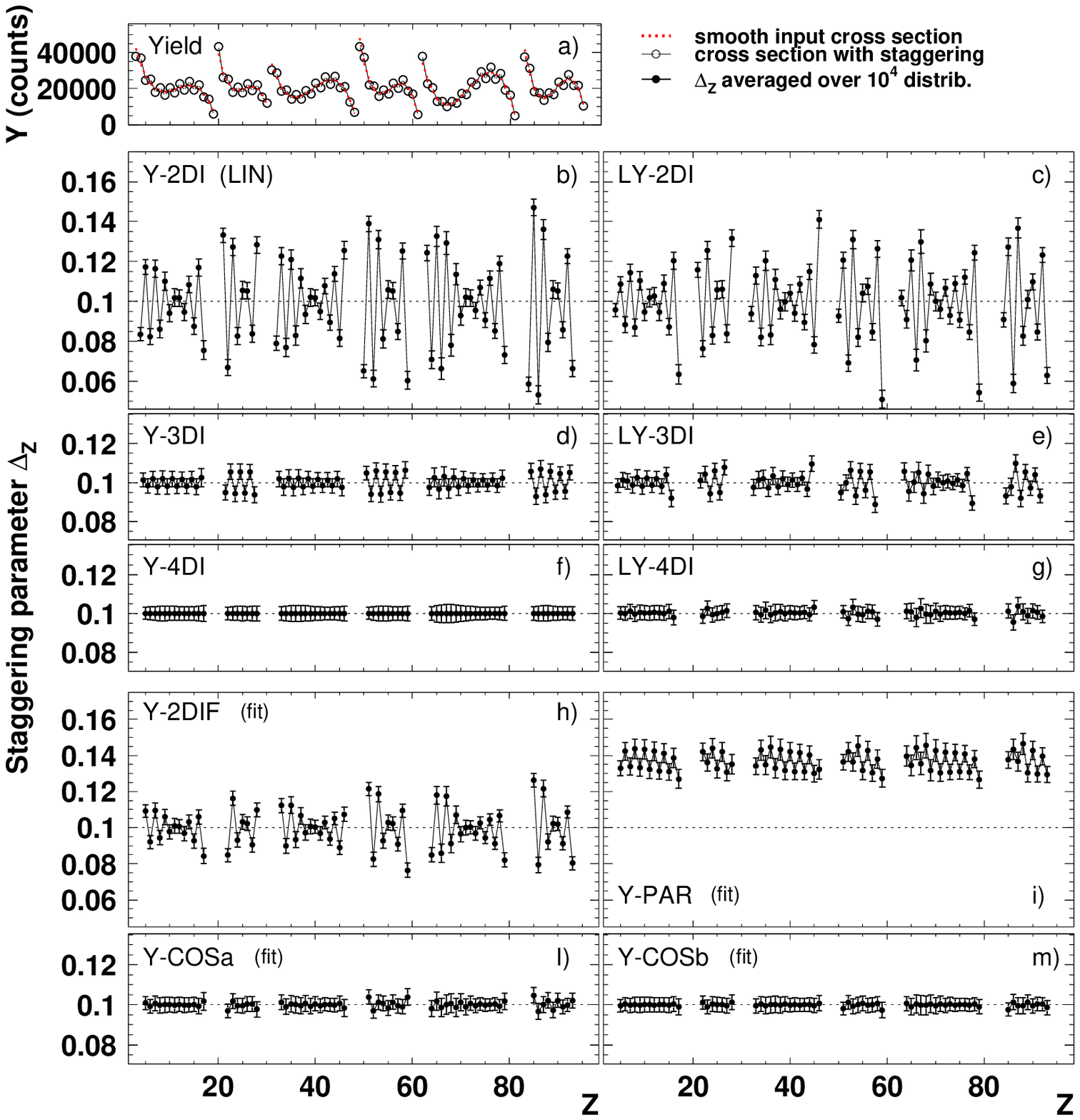}}
  \caption{(color online) 
   Similar to fig. \ref{fig:maxmin}, but for charge 
   distributions with different shapes of the inflection points.
    }
\label{fig:flessi}       
\end{figure*}

In the remaining of this paper we will adopt a constant and representative 
value $\Delta_Z$ = 0.1. 
In fact we played with possible variations of the staggering magnitude and 
of its charge dependence, but nothing noteworthy was found.
On average (or, equivalently, in a single realization with very small 
relative errors) almost all methods do an excellent job when the staggering 
$\Delta_Z$ = 0.1 is applied to a charge distribution that either stays 
constant, or rises or falls in a linear way, whatever the value of its slope.
The only exception concerns the method Y-PAR that, with respect to the other 
methods, amplifies the staggering by about 40\%
(see Appendix B for more details).

\begin{figure*}
\resizebox{0.8\textwidth}{!}{%
  \includegraphics{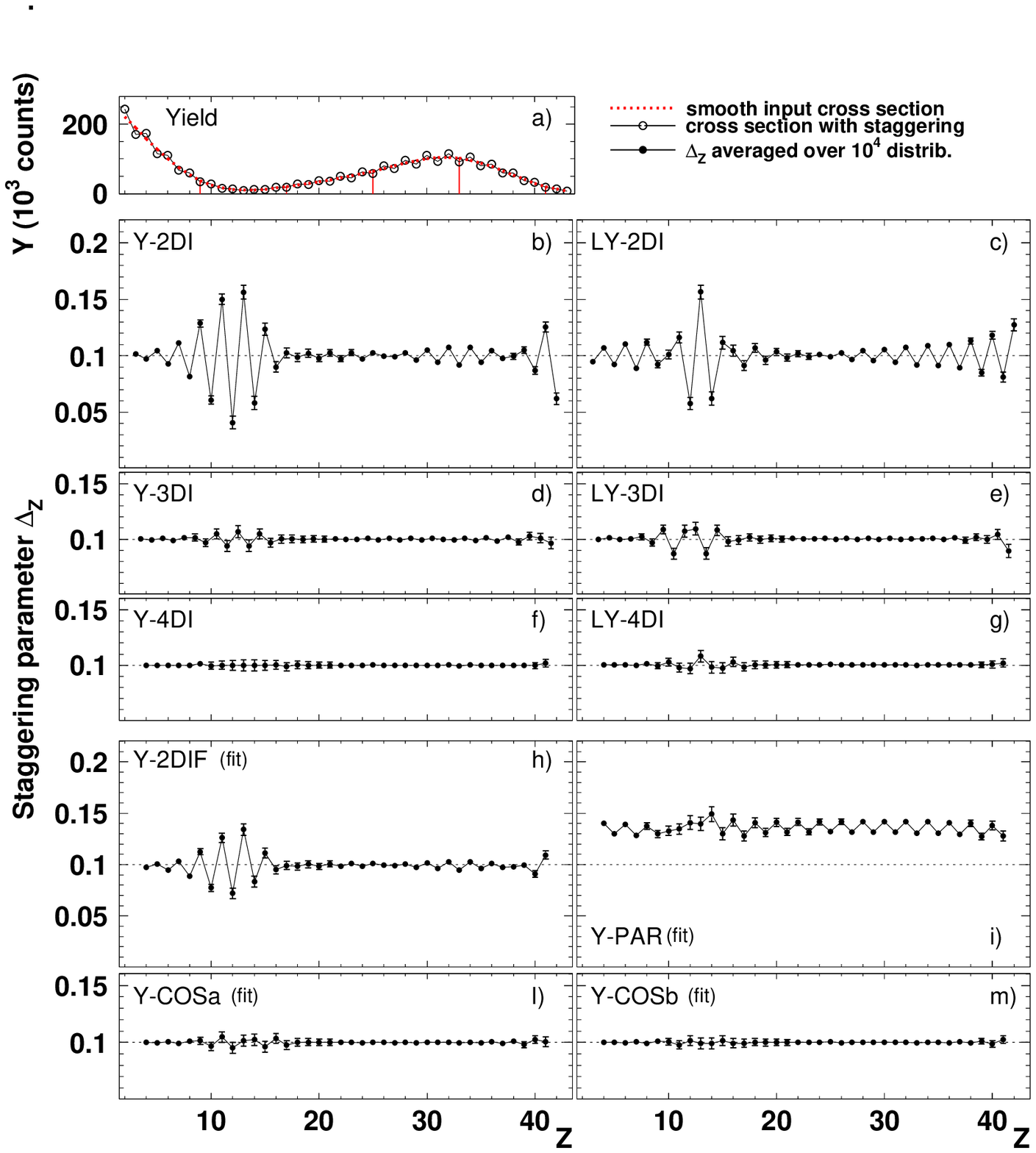}}
  \caption{(color online) Upper panel (a): high-statistic charge distribution
   based on a {\em natural cubic spline\ } (smooth dashed line);
   the thin vertical lines indicate the internal knots (see text).
   Open dots display the effect of a constant staggering $\Delta_Z = 0.1$.
   Panels (b)-(m) present $\Delta_Z$, as obtained with the methods discussed
   in this paper, compared with the nominal value of 0.1 (dotted lines).
   For each $Z$, the error bar indicates the $\pm 1\sigma$ width of the 
   $\Delta_Z$ distribution obtained from $10^4$ simulations.
   }
\label{fig:spline_high}
\end{figure*}

\begin{figure*}
\resizebox{0.8\textwidth}{!}{%
  \includegraphics{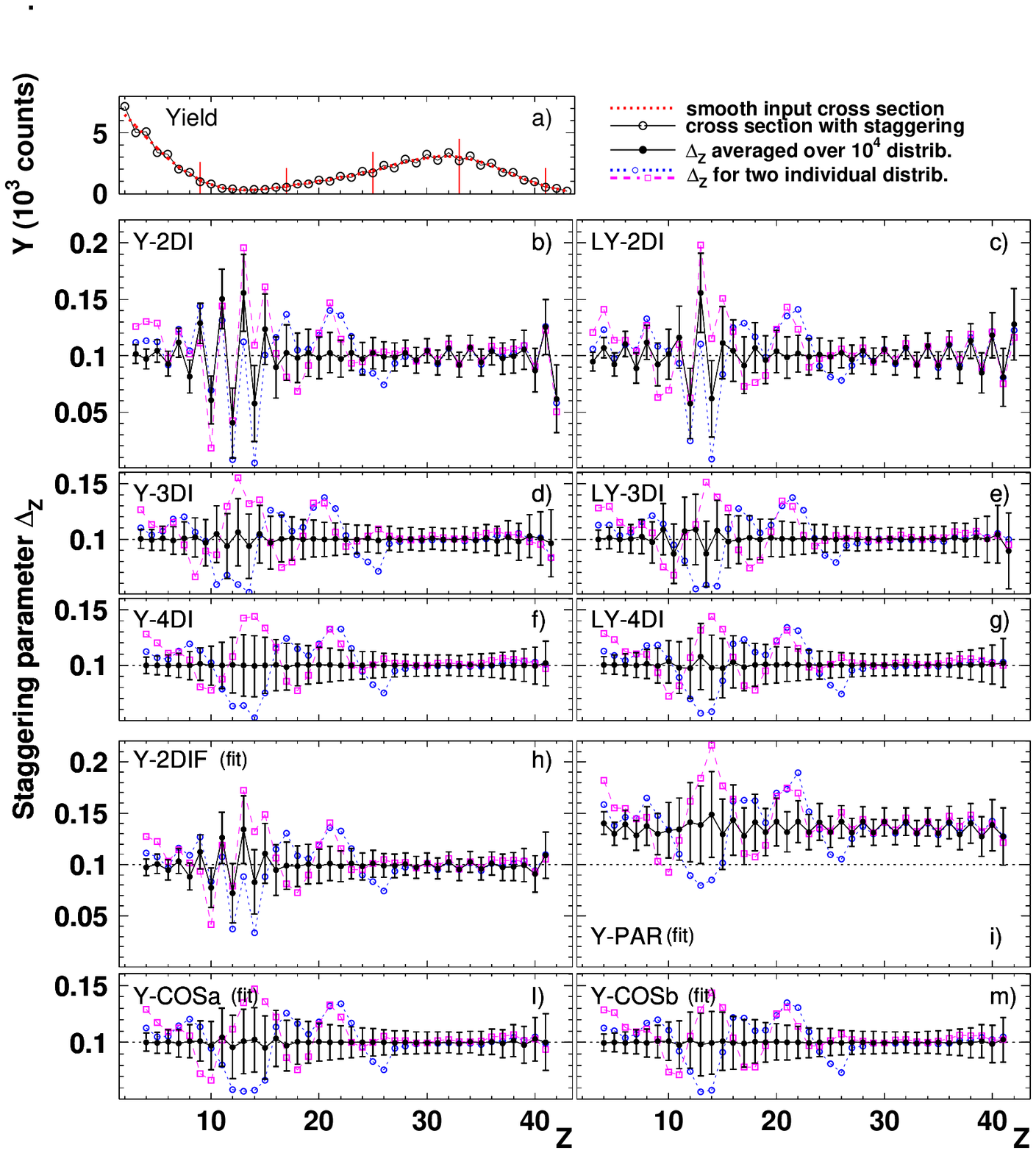}}
  \caption{(color online) Upper panel (a): same charge distribution
   as in fig. \ref{fig:spline_high}, but with low statistic.
   In panels (b)-(m) the average behavior over $10^4$
   realizations (full dots) is the same as in the previous figure,
   only the fluctuations (error bars) expected for a {\em single} realization
   are much larger.
   The open dots/squares (joined by dashed/dotted lines) put into evidence
   the correlated propagation of the statistical fluctuations for two
   randomly selected realizations of the low-statistic charge distribution.
    }
 \label{fig:spline_low}
\end{figure*}

Apart the trivial (and unrealistic) case of a linear charge distribution, 
we now have to investigate how the different methods behave in
regions where the distribution $ Y_\mathrm{sm}(Z)$ is highly not linear 
(namely it presents an appreciable curvature or has an inflection point)
on the basis of some representative cases
that are shown in figs. \ref{fig:maxmin} to \ref{fig:spline_low}.

  The figures are organized as follows.
  The first panel (a) always shows the yield distribution, while the
  remaining panels, from (b) to (m),
  present the result $\langle \Delta_Z \rangle$ obtained with the various
  methods and averaged over $10^4$ replicas of the charge distribution
  (from now on we will omit the $\langle\; \rangle$ for simplicity).
  The results in the first three rows (panels from (b) to (g)) are based 
  on the second, third and fourth finite differences, applied either
  directly to the yields (Y-2DI, Y-3DI and Y-4DI in the left columns),
  or to their logarithms (LY-2DI, LY-3DI and LY-4DI in the right columns).
  The results in the last two rows (panels from (h) to (m)) always refer
  to the methods based on various types of fits to the yield distribution
  (Y-2DIF, Y-PAR, Y-COSa and Y-COSb).
Note that the vertical scales for the methods
Y-2DI and LY-2DI (panels (b) and (c))
and for Y-2DIF and Y-PAR (panels (h) and (i))
span a range about twice as large as that for the other methods.
As already anticipated, the error bars represent the statistical uncertainty 
(at the $\pm 1 \, \sigma$ level) for the analysis of just one 
realization of the distributions shown in the panels (a).

%
Figure \ref{fig:maxmin} shows six examples (three maxima and three minima) 
of charge distributions shaped as vertical parabolas $y(x) = a (x-x_0)^2 + c$.
They have pairwise the same curvature, but with opposite sign
($a = \pm 200$ for the first pair,  $\pm$ 400 for the second 
and $\pm$ 600 for the third one),
while the value in the vertex is always the same, $c = 20000$.
With a staggering parameter of $\Delta_Z$=0.1, the enhancement (reduction) 
of even (odd) charges $Z$ near the vertex is always the same,
i.e. +(-)2000 counts (step 2 in sect. \ref{sec:simul}).

As expected, the simple Y-2DI and LY-2DI, shown in panels (b) and (c), are
not well suited for distributions with an appreciable curvature and indeed
they give the worst results.
For what concerns the remaining finite-difference methods they all behave
rather well.
The method LY-4DI in (g) gives some improvement over LY-3DI, the method by
Tracy shown in (e), especially for the downward parabolas at the two 
extremes of the usable $Z$ range (where the statistic is very low).
However the linear methods Y-3DI (d) and Y-4DI (f) seem to be slightly better.

For the methods based on fits to the data, the addition of two more points
to Y-2DI that is necessary to perform a fit (method Y-2DIF in (h)) certainly
improves the results, especially for the smallest curvatures.
In the Y-PAR method, panel (i), one sees the already mentioned amplification
of about 40\% and the intrinsic difference of even and odd $Z$
(already reduced by treating the errors as explained in Appendix B).
Y-PAR gives similar results for all curvatures (as expected from
its being based on a parabolic fit), but it certainly does not represent 
a real improvement over Y-2DIF.
Much better results are definitely obtained by the two new methods,
Y-COSa and Y-COSb (based on the fit with a properly oscillating function).

%
The situation is in general even worse for inflection points, which make 
more trouble for nearly all methods.
In fig. \ref{fig:flessi}(a) we present six charge distributions 
that differ either in the slope at the inflection point or in the distance 
between the minimum and the maximum. 
The distributions are obtained from cubic functions without quadratic term,
$Y(Z) = a\, (Z-Z_0)^3 + b\, (Z-Z_0) +\,c$, all with the same statistic 
at the inflection point ($c = 20000$). 
The slope at the inflection point is $b$ = 1000, 2000 and 3000 for the first, 
second and third pair of distributions, respectively, while the cubic 
coefficient is $a$ = $-$50 for the first and $-$150 for the second curve 
of each pair.

The lower panels of fig. \ref{fig:flessi}, from (b) to (m), present the
average results for the different methods.
The methods Y-2DI and LY-2DI of panels (b) and (c) are certainly not acceptable,
because they give systematic deviations up to $\pm 0.05$ with respect to the
nominal value of $\Delta_Z$=0.1.
Of the remaining finite-difference methods, Y-4DI and LY-4DI of panels (f)
and (g) are now very remarkable improvements over Y-3DI (d) and LY-3DI by Tracy (e),
because they reduce the deviations from the nominal value down to less than $\pm$0.005.
This does not come totally unexpected, because methods based on the fourth differences
should be able to remove smooth terms up to the third order (see Appendix A).

The fit method Y-2DIF of panel (h) represents again
an appreciable improvement over the linear method Y-2DI (b), as it cuts
the systematic deviations down by about a factor of two, 
but the method Y-PAR of panel (i), in spite of its amplification and 
systematic even-odd effect, performs even better than Y-2DIF.
What comes rather unexpected is that the methods Y-COSa (l) and especially Y-COSb (m)
perform rather well, in a way comparable with Y-4DI (f) and LY-4DI (g).
Altogether, the best results among all those displayed in fig. \ref{fig:flessi}
are obtained with the methods Y-4DI and Y-COSb.

The high statistic adopted for the smooth distribution
$Y_\mathrm{sm}(Z)$ (with relative errors lower than $\pm$1\% in the non-linear
regions) simply makes the presentation of the data in fig. \ref{fig:maxmin}
and \ref{fig:flessi} more readable by reducing the error bars, but obviously
does {\em not} appreciably change the response of the methods,
averaged over many distributions~\footnote{
  For example, one sees from eqs. \eqref{eq:3di}, \eqref{eq:4di}
  or \eqref{eq:Delta} that a common factor $k$ multiplying all yields cancels
  out in the average $\langle \Delta_Z \rangle$; and it cancels out also
  in $\langle D_Z \rangle$, when
  $\langle \ln(k\,Y) \rangle= \ln(k)+\langle \ln(Y) \rangle$ 
  is inserted in eqs. \eqref{eq:l3di} or \eqref{eq:l4di}.}.
Of course, for a single realization (or experiment) a large statistic of 
the charge distribution remains of paramount importance.

A second point worth noting is that the deviations from the nominal value 
displayed in fig. \ref{fig:maxmin} and \ref{fig:flessi} are characteristic
of each method and of the specific non-linearities of the
smooth function $Y_\mathrm{sm}$.
The simulation also shows that in general the deviations from the nominal
value of the staggering parameter are very weakly dependent on the chosen
value of $\Delta_Z$.
For methods based on the yield logarithms, one can even demonstrate
that these deviations do not depend on it.
Considering for example the case of LY-3DI, if $L(Z)$ has the form of
eq. \eqref{eq:Lcubic} with a cubic coefficient $a$, then eq. \eqref{eq:l3di}
gives $D_Z +\tfrac{3}{4} a$ (instead of $D_Z$),
thus with a constant deviation from the nominal value that amounts to
$\tfrac{3}{4} a$ and hence does not depend on $D_Z$.
Therefore, the choice of a {\em good} method is important,
especially when the average staggering effects are small.

We tested several other shapes of the charge distribution, finding that in 
all cases, irrespective of the specific shape, the ranking of the various 
methods remains similar to that found in 
figs. \ref{fig:maxmin} and \ref{fig:flessi}.

We show just one case in figs. \ref{fig:spline_high} and \ref{fig:spline_low},
where a more realistic charge distribution -- open dots in panels (a) -- mimics 
the typical features found in semi-peripheral collisions of
intermediate-energy heavy-ion reactions:
with increasing $Z$ there is a very rapid drop of the cross section from 
the lightest complex fragments with $Z \approx 4$ down to a wide valley 
of intermediate mass fragments (IMF) around $Z \approx$ 10 -- 15, 
followed by a slow rise towards heavier ones (which may represent 
either fission-like or projectile-like fragments) until it finally fades away.

This shape was obtained 
with a {\em natural cubic spline} passing through seven equally spaced knots 
(the five internal ones are indicated by thin vertical lines in the figures):
it is a piece-wise interpolation with $3^\mathrm{rd}$ degree polynomials
that is continuous in the joining knots up to the second derivative.
The adjective {\em natural} indicates the additional condition 
that the second derivative is zero in the two external knots.

In fig. \ref{fig:spline_high}
the statistic for each of the $10^4$ generated replicas is very large:
in the IMF valley it corresponds to at least 
ten thousand counts for each charge $Z$.
The lower panels, from (b) to (m), illustrate again the estimates of  
$\Delta_Z$ obtained with the various methods
already shown in the previous figures.
The full dots represent the average obtained over $10^4$ simulations,
while the error bars, often smaller than the symbol sizes,
give the widths of the distributions for each $Z$.
Again dashed lines at $\Delta_Z = 0.1$ indicate the nominal value.

Also here we observe an evident failure of Y-2DI, LY-2DI and Y-2DIF
in the minimum of the IMF valley and in the final tail of the distribution.
Of all finite-difference methods,
those based on the fourth differences perform better
(Y-4DI even slightly better than LY-4DI), but comparably good results
are obtained with the fitting methods Y-COS and 
particularly with the second one, Y-COSb.

Concerning the finite difference methods, one might think that fragment
cross sections that vary over many orders of magnitude are better
studied with the logarithmic n-point differences.
The present study does not lend much support to this idea, as the results of
linear and logarithmic methods presented in figures from \ref{fig:maxmin}
to \ref{fig:spline_high} do not differ too much from each other.
For the second-difference results of panels (b) and (c), one might have
a slight preference for the logarithmic methods, but the choice is hard.
The situation for the third and fourth differences
(panels (h) vs. (i) and (i) vs. (m), respectively) seems to be slightly
in favour of the methods that use directly the yields.
In fact, the logarithmic methods seem to suffer more in regions where
the yield is lower, see e.g. the IMF valley of fig. \ref{fig:spline_high},
or the sides of the downward parabolas in fig. \ref{fig:maxmin}.

In order to stress the importance and the effects of the statistic,
fig. \ref{fig:spline_low} presents results for the same charge distribution
of fig. \ref{fig:spline_high}, but in the case of low statistic:
about 300 counts per charge $Z$ in the IMF valley.
The average behavior is substantially the same as in the previous figure.
In fact, for each of the presented methods, the full dots
are practically the same in both figures, their differences being much
smaller than the symbol sizes.
The main difference concerns
the $\pm 1 \sigma$ fluctuations (error bars) expected for a single realization
(or measurement), which are of course much larger in fig. \ref{fig:spline_low}.

In addition, two examples of the results obtained for two individual
random realizations of the charge distribution are shown by the open
dots/squares, joined by dashed/dotted lines in the panels
(b)-(m) of fig. \ref{fig:spline_low}.
Here the effects of the statistical fluctuations are clearly visible 
for all methods and reveal marked deviations from the average behavior,
with several points located outside the error bars.
What is even worse is the fact that the fluctuations give rise to 
structures involving several neighboring points, a consequence of their 
correlation (see next section).  
The location and the amplitude of these nonphysical structures present large 
random variations from one single realization to the other, being sometimes 
in-phase and sometimes out-of-phase in the two realizations.
This means that in a single experiment,
unless the independent relative errors, including the statistical ones,
are very small (below the 1\% level), the physical 
significance of structures in the staggering parameter $\Delta_Z$ should 
be treated with great caution.

\section{Correlations}
The need of estimating the smooth behavior of the charge distribution from a few
data points, naturally introduces short-range autocorrelations among the staggering
parameters $\Delta_Z$ pertaining to neighboring charges $Z$.
The degree of autocorrelation is estimated with the usual definition
\begin{equation}
R(Z',Z'') = \frac{\langle \;
           (\Delta_{Z'} - \overline{\Delta_{Z'}}) \,
           (\Delta_{Z''} - \overline{\Delta_{Z''}}) \;\rangle}
             {\sigma_{\Delta_{Z'}} \, \sigma_{\Delta_{Z''}}} 
 \makebox[5mm][c]{}         \label{eq:correl}
\end{equation}
where the overlined $\overline{\Delta_Z}$'s are the mean values of the 
staggering parameters for two different charges $Z'$ and $Z''$,
the brackets $\langle \; \rangle$ indicate the expectation value of the 
product of the deviations of the two $\Delta_Z$ from their mean values
and the $\sigma$'s in the denominator are their standard deviations.
By construction $R(Z',Z'')$ lies in the range $[-1,1]$, the two extremes 
indicating perfect correlation or anticorrelation, respectively
(while 0 indicates uncorrelated data).

The correlation factor $R$ cannot be estimated from a single experiment, 
but it is easily obtained in a simulation where many replicas of a certain 
charge distribution are generated, differing only in the fluctuations.
As expected, fluctuations introduce a correlation that is positive for 
two nearby located $Z'$ and $Z''$ values and 0 for distant ones.

The correlation factor is practically independent of the
specific shape of the analyzed charge distributions.
For a flat distribution with a constant $\Delta_Z$ of 0.1, 
the region of positively correlated neighboring $Z$ values 
is obviously at most three charge units wide for the methods
Y-2DI and LY-2DI  (with FWHM values of 2.37$\pm$0.03) and increases to
four charge units for Y-3DI and LY-3DI (with FWHM values of about 2.84)
and to five charge units for all remaining methods
(with a common FWHM of 3.24).
In conclusion, structures possibly found in the staggering analysis
that are narrower than the above numbers are highly suspicious and
probably they should not be given much physical significance.

\section{Conclusions}

The odd-even staggering is a widespread phenomenon that
usually consists in a relative enhancement of the yields of nuclei with 
even values of proton (neutron) number $Z$ ($N$) with respect to
the neighboring nuclei with odd values.
All methods are based on the idea, first proposed by Amiel and 
Feldstein \cite{Amiel74,Amiel75}, that the measured yield $Y(Z)$ can be 
factorized into the product of a smooth yield $Y_\mathrm{sm}(Z)$ 
times a factor $(1 \pm \Delta_Z)$, where the sign `$+$' holds for even 
charges and the sign `$-$' for odd ones.

This is an assumption, widely accepted and probably justifiable on the 
basis of theoretical arguments, but it remains an assumption and 
one should definitely remain aware of that.

We have reviewed some old methods and we propose some new methods of 
extracting the staggering parameter $\Delta_Z$ (or $D_Z$)
from the experimental charge distributions.
Some methods first estimate the underlying smooth behavior of the yield  
$\mathcal{Y}(Z)$ from the data themselves, to derive then the staggering 
from the difference between the measurement and the smoothed yield, 
while others deliver a direct estimate of the staggering parameter.
Some methods use a fit to the data, others take advantage
of the finite-difference mathematical formalism.

Almost all methods perform equally well if the charge distribution 
has a predominant linear behavior, while they greatly differ from 
each other in the presence of strongly non-linear regions
in the charge distribution.
So steepness is not an issue, not even for fragment yields.
From the present study, using logarithmic n-point differences
for fragment yields is just an option, not a real necessity.

All methods are extremely sensitive to the fluctuations that may affect 
the yields of neighboring $Z$ in the experimental data, giving
origin to spurious structures without much physical meaning.
Therefore for this kind of studies it is of paramount importance to acquire 
good data with very large statistic and very small independent relative errors, 
so that the relative random fluctuations 
for each measured $Z$ are of the order of about 1\% or better: only
when this condition is fulfilled, one can really
appreciate the different quality of the various methods.
On the contrary, if the fluctuations are sizably larger, spurious effects 
appear, and it makes no much sense to choose one method rather than another; 
the results will be anyhow poor and hard to interpret in a sensible way.

We find that linear interpolations (Y-2DI and LY-2DI) that make use of
three neighboring points give the worst results among all examined methods.
Extending the linear interpolation with a fit that uses five points (Y-2DIF)
brings only some moderate improvement.
The recently proposed fit of a simple parabola over five points (Y-PAR)
does not represent a real improvement over Y-2DIF.
The method Y-PAR presents the peculiarity that it amplifies the signal, but
it needs a careful treatment of the errors and even then it does not produce a 
really smooth estimate of the distribution without staggering, $\mathcal{Y}(Z)$.

Very good results are given by the newly proposed method (Y-COS) that 
uses a five-point fit with a properly oscillating function.
As it implies a non-linear fit, a reliable recursive procedure has to 
be used, such as the well known \textsc{Minuit} code \cite{minuit}
developed in \textsc{CERN}. 
Using directly the coefficient of the oscillating term (method Y-COSb) 
as delivered by the fit seems to be one of the best ways
to estimate the staggering parameter $\Delta_Z$.

A completely different approach is the oldest one by Tracy,
based on the third (LY-3DI) finite differences over four values
that are the logarithms of the fragment yields.
At variance with the previous methods, it delivers
directly an estimate of the staggering parameter, although with a
slightly different parametrization first proposed by Wahl \cite{Wahl80}.
We show that a similar method (Y-3DI) could be applied to the
fragment yields, using the parametrization by Amiel \cite{Amiel74,Amiel75}.
In both cases the staggering value, that is usually attributed to the center of
the considered interval, corresponds to a half-integer $Z$.
These two methods give good results, but not as good as those of Y-COSb.

Further exploiting the Finite Difference formalism,
in this paper we propose two more methods, Y-4DI and LY-4DI, that use
the fourth finite differences over five points.
They give somewhat better results than the original LY-3DI because they are
insensitive to possible cubic components in the smooth charge distribution, 
and probably also because they use one more point and hence they are slightly 
less sensitive to fluctuations.
Moreover, one practical (and aesthetic) advantage is that the results can be 
attributed to the integer $Z$ values at the center of the five-point intervals.

Concluding, to avoid the appearance in $\Delta_Z(Z)$ of structures without 
much physical significance 
i) one should collect data with a very large statistic;
ii) all other uncertainties that may affect the yields of neighboring $Z$ 
in an independent way should also be carefully estimated and reduced;
iii) once the previous conditions are met, the methods which seem less 
sensitive to non-linearities of the underlying smooth distribution 
$Y_\mathrm{sm}(Z)$ are the methods
Y-4DI and Y-COSb, closely followed by LY-4DI and Y-COSa.


\section*{Appendix A} \label{sec:appA}

\setcounter{equation}{0}
\renewcommand{\theequation}{A\arabic{equation}}

The Finite Difference Calculus (see e.g. \cite{Boole1860,Jordan1950,Spiegel1970})
is a widely used mathematical formalism dealing with finite
increments of the independent variable(s) of mathematical functions.
In particular, here we consider a function $y(x)$ whose values are known
for a series of equidistant values of the independent variable $x$;
such is the case of the fragment yield $Y(Z)$ that is defined
only for integer values of $Z$.

The known values of the function form a sequence $y_k$, where the subscript
$k=0, 1, 2 \ldots$ indicates the number of equal increments with respect to
a given reference value $x_0$
(in the general case $k$ can take negative integer values too).
From the sequence $y_k$ one can build the first difference
$\mathcal{D}y_k \equiv (y_{k+1}-y_k)$, and all differences of higher
order~\footnote{
 In mathematical textbooks the difference and derivative operators of order
 $n$ are usually represented with $\Delta^n$ and $D^n$, respectively.
 Unfortunately, the same symbols are commonly used for the staggering
 parameters of Amiel/Wahl.
 For the sake of clarity, here we use $\mathcal{D}^{n}$ for the
 finite difference operator.}
\begin{equation}
  \mathcal{D}^{n}y_k \equiv (\mathcal{D}^{n-1}y_{k+1}-\mathcal{D}^{n-1}y_k)
         =  \sum_{i=0}^{n} (-1)^i \binom{n}{i} y_{k+i}      \label{eq:Dn}
\end{equation}
\noindent
where $\binom{n}{i}$ are the binomial coefficients.

The property relevant for the present paper is the following:
if the sequence $y_k$ is described by a polynomial of degree $n$-1
in the index $k$, namely $y_k= (\sum_{i=0}^{n-1} c_i k^i)$, then the differences of
order $n$ (built from $n$+1 values of $y_k$) are identically zero
(and so is for those of higher order too)~\footnote{
  If $y_k$ is generated by a function that is not a polynomial,
  then the finite difference operator $\mathcal{D}^{n}$ can be used
  to remove all smooth terms in the Taylor series up to the order $n$-1.}.

If one further {\em assumes} that the sequence $y_k$, described by a polynomial
of degree $n$-1, is perturbed by a staggering parameter $\Delta$ {\em \`a la} Amiel
\begin{equation}
  y_k =  \Big( \sum_{i=0}^{n-1} c_i\, k^i \Big) \Big(1 + (-1)^k \Delta \Big),
                                                       \label{eq:p_amiel_1}
\end{equation}
then a little algebra gives
\begin{equation}
  \mathcal{D}^{n}y_k
      \equiv  (-1)^k \Delta\; \Big( \sum_{i=0}^{n} \binom{n}{i} y_{k+i} \Big)
                                                       \label{eq:pol_amiel_2}
\end{equation}
\noindent 
and hence the ($n$+1)-point formula:
\begin{equation}
  \Delta_{k+n/2} =  \left.  (-1)^k \; \mathcal{D}^{n}y_k \; \right/
     \Big( \sum_{i=0}^{n} \binom{n}{i} y_{k+i} \Big)         \label{eq:pol_amiel_3}
\end{equation}
\noindent
where we have added to $\Delta$ the subscript $k+n/2$ indicating the central
value of the used interval $[y_k, y_{k+n}]$.

Assuming that the polynomial of degree $n$ describes the {\em logarithm}
of the sequence $y_k$, it results more convenient to use the staggering
model {\em \`a la} Wahl and according to eq. \eqref{eq:wahlog} one can write
\begin{equation}
  \ln(y_k) =  \Big( \sum_{i=0}^{n-1} c_i\, k^i \Big) \; + \; (-1)^k \; D,
                                                       \label{eq:pol_wahl_1}
\end{equation}
which, with a little algebra, gives
\begin{equation}
  \mathcal{D}^{n}y_k
     \equiv  (2^n) \; \cdot (-1)^k \; D\;             \label{eq:pol_wahl_2}
\end{equation}
\noindent 
and again an ($n$+1)-point formula for $D$:
\begin{equation}
  D_{k+n/2} = (-1)^k \; \frac{1}{2^n} \;  \mathcal{D}^{n}y_k.
                                                        \label{eq:pol_wahl_3}
\end{equation}
\noindent
where again the subscript $k\!+\!n/2$ indicates the central value
of the used interval.

Identifying the fragment yield $Y(Z)$ with $y_k$ (and with a trivial index
redefinition) these equations generate all the methods of
Sect. \ref{sec:methods} that do not make use of a fit to the yield data.
For example, from eq. \eqref{eq:pol_wahl_3} with $n\,=\,3$,
using the definition of eq. \eqref{eq:Dn},
one obtains the method LY-3DI of eq. \eqref{eq:Ltracy},
first proposed long time ago by Tracy \cite{Tracy72};
and with $n\,=\,2$ (4) one obtains the method LY-2DI (LY-4DI).
Similarly, from eq. \eqref{eq:pol_amiel_3} with $n\,=\,2,$ 3 and 4, one
obtains the methods Y-2DI, Y-3DI and Y-4DI, respectively.
\vspace{3mm}

\section*{Appendix B} \label{sec:appB}

\begin{figure}[b]
 \resizebox{0.50\textwidth}{!}{%
  \includegraphics{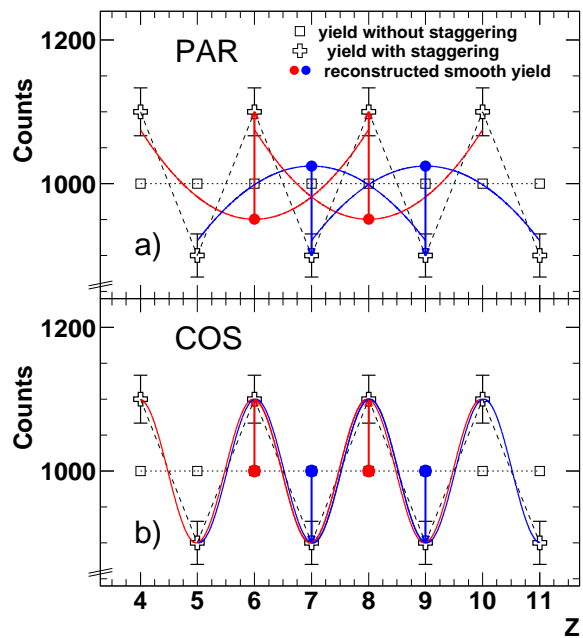}}
  \caption{(color online) Effect of applying a {\em weighted least square fit}
   in the Y-PAR (a) and Y-COS (b) methods, with the weights taken from the 
   statistics of the data.
   The original distribution (squares) is flat with a statistic of 1000 counts;
   a $\Delta_Z$=0.1 staggering effect (crosses) is superimposed on it.
   In the Y-PAR method the differing statistical errors for even and odd $Z$ 
   induce an artifact that estimates a larger staggering for even $Z$ than 
   for odd ones (see text).
    }
 \label{fig:app}       
\end{figure}

\begin{figure}[t]
 \resizebox{0.50\textwidth}{!}{%
  \includegraphics{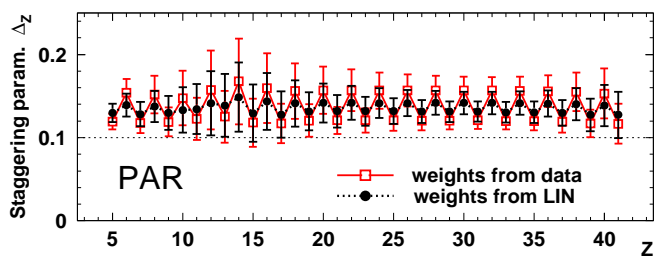}}
  \caption{(color online) 
   Staggering parameter $\Delta_Z$ for the charge distribution
   of fig. \ref{fig:spline_low}(a) obtained with the method Y-PAR, using the
   {\em weighted least square fit} with statistical errors estimated
   directly from the experimental data (open squares), or by first applying
   the LIN/Y-2DI linear method (full dots).
    }
 \label{fig:compN}       
\end{figure}

With respect to the others, the method Y-PAR amplifies the estimated $\Delta_Z$
by about 40\% above the nominal value.
This has to be taken into account when quantitatively comparing the results of
this method with those of the others.
A second characteristic intrinsic in the Y-PAR method 
is that this amplification is not the same for even and odd charges.
It is somewhat larger for even charges (usually enhanced by
staggering) than for odd ones, so that even in case of constant input values
of the yield $Y_\mathrm{sm}(Z)$ and of the parameter $\Delta_Z$, one does not
obtain a really flat $\Delta_Z$ distribution, even for an input value of
$\Delta_Z$=0 (see full dots in panels of row (d) in fig. \ref{fig:zero}).

This effect is connected with the weights used in the 
{\em weighted least squares fit} procedure.
When the smoothed yield $\mathcal{Y}(Z)$ is deduced by means of a fit
(like in the Y-PAR and Y-COS methods), some care must be put in the treatment 
of the errors on the measured yields $Y(Z)$.
In fact the {\em weighted least square fit method} may introduce a subtle 
asymmetry between even and odd $Z$ values when the weights mainly
depend on the number of counts measured for each $Z$.
In this case, even $Z$ yields (which are enhanced) have systematically somewhat
larger errors (and hence smaller weights) than odd $Z$ yields.

This causes an artifact in the Y-PAR method because the parabola is not 
the appropriate function to fit the five `zigzagging' experimental points.
This is shown in fig.~\ref{fig:app}(a), where a constant yield of 1000 counts 
with $\Delta_Z$=0.1 is assumed, leading to statistical weights of about
$1/ \sqrt{1100}$ and $1/ \sqrt{900}$ for the even and odd $Z$, respectively.
With their larger weights, the three reduced yields (lower crosses) of an 
odd $Z$ are more effective in shifting the parabola downward
than the three enhanced yields (upper crosses) of an even $Z$
in shifting it upward, due to their smaller weights: 
as a consequence, even $Z$ values appear to have a larger staggering 
than odd ones.
We stress that this effect {\em cannot} be cured by increasing the
statistic of the measurement.
It is worth noting that, on the contrary, the fits of the Y-COS method
shown in fig. \ref{fig:app}(b) are not affected by this problem, because this
method uses a properly oscillating fit function that well reproduces the data.

The open squares of fig. \ref{fig:compN} show the effect on $\Delta_Z$ when 
the errors in Y-PAR are estimated directly from the data, in case of the charge 
distribution already presented in fig. \ref{fig:spline_low}(a).
This artifact may be cured either by recurring to the 
{\em unweighted least square method} 
(which is however unsatisfactory when the yields strongly vary with $Z$)
or performing appropriate averages of the weights before fitting the data.
In this paper we always used as statistical weights for the Y-PAR method 
the results of a preceding analysis performed with the Y-2DI linear method. 
This artifice proved essential to obtain results (full dots in fig. 
\ref{fig:compN}) that could be meaningfully compared with the other methods.\\

We wish to warmly thank Prof. P.R. Maurenzig for continuous support,
useful suggestions and very fruitful discussions.

\vspace{10mm}


\bibliography{metodiNR_rev}

\begin{thebibliography}{10}

\bibitem{Runnalls69}
N.~G. Runnalls, D.~E. Troutner, and Robert~L. Ferguson.
\newblock {\em Phys. Rev.}, 179:1188, 1969.

\bibitem{Tracy72}
B.~L. Tracy, J.~Chaumont, R.~Klapisch, J.~M. Nitschke, A.~M. Poskanzer,
  E.~Roeckl, and C.~Thibault.
\newblock {\em Phys. Rev. C}, 5:222, 1972.

\bibitem{Clerc75}
H.-G. Clerc, W.~Lang, H.~Wohlfarth, K.-H. Schmidt, H.~Schrader, K.~E.
  Pferdek\"amper, and R.~Jungmann.
\newblock {\em Z. Phys. A}, 274:203, 1975.

\bibitem{Siegert76}
G.~Siegert, H.~Wollnik, J.~Greif, R.~Decker, G.~Fiedler, and B.~Pfeiffer.
\newblock {\em Phys. Rev. C}, 14:1864, 1976.

\bibitem{Tsekhanovich99}
I.~Tsekhanovich, H.O. Denschlagb, M.~Davib, Z.~B{\"u}y{\"u}kmumcuc,
  M.~W{\"o}stheinrichd, F.~G{\"o}nnenweind, S.~Oberstedte, and H.R. Fauste.
\newblock {\em Nucl. Phys. A}, 658:217, 1999.

\bibitem{Schmidt01}
K.-H. Schmidt, J.~Benlliure, and A.~R. Junghans.
\newblock {\em Nucl. Phys. A}, 693:169, 2001.

\bibitem{Tsekhanovich04}
I.~Tsekhanovich, N.~Varapai, V.~Rubchenya, D.~Rochman, G.~S. Simpson,
  V.~Sokolov, G.~Fioni, and {Ilham Al Mahamid}.
\newblock {\em Phys. Rev. C}, 70:044610, 2004.

\bibitem{Naik07}
H.~Naik, S.~P. Dange, and A.~V.~R. Reddy.
\newblock {\em Nucl. Phys. A}, 781:1, 2007.

\bibitem{Rejmund00}
F.~Rejmund, A.~V. Ignatyuk, A.~R. Junghans, and K.-H. Schmidt.
\newblock {\em Nucl. Phys. A}, 678:215, 2000.

\bibitem{Wahl62}
A.~C. Wahl, R.~L. Ferguson, D.~R. Nethaway, D.~E. Troutner, and K.~Wolfsberg.
\newblock {\em Phys. Rev.}, 126:1112, 1962.

\bibitem{Gonnenwein92}
F.~G\"onnenwein.
\newblock {\em Nucl. Instrum. Methods A}, 316:405, 1992.

\bibitem{Amiel74}
S.~Amiel and H.~Feldstein.
\newblock In {\em Third International Symposium on Physics and Chemistry of
  Fission}, volume~II, page~65. IAEA, Vienna, 1974.

\bibitem{Amiel75}
S.~Amiel and H.~Feldstein.
\newblock {\em Phys. Rev. C}, 11:845, 1975.

\bibitem{Wahl80}
A.~C. Wahl.
\newblock {\em J. Radioanalytical Chemistry}, 55:111, 1980.

\bibitem{Poskanzer71}
A.~M. Poskanzer, G.~W. Butler, and E.~K. Hyde.
\newblock {\em Phys. Rev. C}, 3:882, 1971.

\bibitem{Webber90}
W.~R. Webber, J.~C. Kish, and D.~A. Schrier.
\newblock {\em Phys. Rev. C}, 41:533, 1990.

\bibitem{Knott96}
C.~N. Knott, S.~Albergo, Z.~Caccia, C.-X. Chen, S.~Costa, H.~J. Crawford,
  M.~Cronqvist, J.~Engelage, P.~Ferrando, and R.~Fonte {\em et al.,}.
\newblock {\em Phys. Rev. C}, 53:347, 1996.

\bibitem{Zeitlin97}
C.~Zeitlin, L.~Heilbronn, J.~Miller, S.~E. Rademacher, T.~Borak, T.~R. Carter,
  K.~A. Frankel, W.~Schimmerling, and C.~E. Stronach.
\newblock {\em Phys. Rev. C}, 56:388, 1997.

\bibitem{Ricciardi04}
M.~V. Ricciardi, A.~V. Ignatyuk, A.~Keli\v{c}, P.~Napolitani, F.~Rejmund, K.-H.
  Schmidt, and O.~Yordanov.
\newblock {\em Nucl. Phys. A}, 733:299, 2004.

\bibitem{Napolitani07}
P.~Napolitani, K.-H. Schmidt, L.~Tassan-Got, P.~Armbruster, T.~Enqvist,
  A.~Heinz, V.~Henzl, D.~Henzlova, A.~Keli\'{c}, and R.~{Pleska\v{c} \em et
  al.,}.
\newblock {\em Phys. Rev. C}, 76:064609, 2007.

\bibitem{Napolitani11}
P.~Napolitani, K.-H. Schmidt, and L.~Tassan-Got.
\newblock {\em J. Phys. G}, 38:115006, 2011.

\bibitem{Yang99}
L.~B. Yang, E.~Norbeck, W.~A. Friedman, O.~Bjarki, F.~D. Ingram, R.~A. Lacey,
  D.~J. Magestro, M.~L. Miller, A.~Nadasen, and R.~Pak {\em et al.,}.
\newblock {\em Phys. Rev. C}, 60:041602, 1999.

\bibitem{Winchester00}
E.~M. Winchester, J.~A. Winger, R.~Laforest, E.~Martin, E.~Ramakrishnan, D.~J.
  Rowland, A.~Ruangma, S.~J. Yennello, G.~D. Westfall, {A. Vander Molen}, and
  E.~Norbeck.
\newblock {\em Phys. Rev. C}, 63:014601, 2000.

\bibitem{Geraci04}
E.~Geraci, M.~Bruno, M.~{D'Agostino}, E.~{De Filippo}, A.~Pagano, G.~Vannini,
  M.~Alderighi, A.~Anzalone, and L.~Auditore {\em et al.,}.
\newblock {\em Nucl. Phys. A}, 732:173, 2004.

\bibitem{DAgostino11}
M.~{D'Agostino}, M.~Bruno, F.~Gulminelli, L.~Morelli, G.~Baiocco, L.~Bardelli,
  S.~Barlini, F.~Cannata, G.~Casini, and E.~Geraci {\em et al.,}.
\newblock {\em Nucl. Phys. A}, 861:47, 2011.

\bibitem{Ademard11}
G.~Ademard, J.~P. Wieleczko, J.~Gomez del Campo, M.~La Commara, E.~Bonnet,
  M.~Vigilante, A.~Chbihi, J.~D. Frankland, E.~Rosato, and G.~Spadaccini {\em
  et al.,}.
\newblock {\em Phys. Rev. C}, 83:054619, 2011.

\bibitem{Lombardo11}
I.~Lombardo, C.~Agodi, F.~Amorini, A.~Anzalone, L.~Auditore, I.~Berceanu,
  G.~Cardella, S.~Cavallaro, M.~B. Chatterjee, and E.~De~Filippo {\em et al.,}.
\newblock {\em Phys. Rev. C}, 84:024613, 2011.

\bibitem{DAgostino12}
M.~{D'Agostino}, M.~Bruno, F.~Gulminelli, L.~Morelli, G.~Baiocco, L.~Bardelli,
  S.~Barlini, F.~Cannata, G.~Casini, and E.~Geraci {\em et al.,}.
\newblock {\em Nucl. Phys. A}, 875:139, 2012.

\bibitem{Casini12}
G.~Casini, S.~Piantelli, P.~R. Maurenzig, A.~Olmi, L.~Bardelli, S.~Barlini,
  M.~Benelli, M.~Bini, M.~Calviani, and P.~Marini {\em et al.,}.
\newblock {\em Phys. Rev. C}, 86:011602(R), 2012.

\bibitem{Piantelli13}
S.~Piantelli, G.~Casini, P.~R. Maurenzig, A.~Olmi, S.~Barlini, M.~Bini,
  S.~Carboni, G.~Pasquali, G.~Poggi, A.~A. Stefanini, and S.~Valdr\`e {\em et
  al.,}.
\newblock {\em Phys. Rev. C}, 88:064607, 2013.

\bibitem{Colonna05}
M.~Colonna and F.~Matera.
\newblock {\em Phys. Rev. C}, 71:064605, 2005.

\bibitem{Raduta05}
Ad.~R. Raduta and F.~Gulminelli.
\newblock {\em Phys. Rev. C}, 75:044605, 2007.

\bibitem{Huang10}
M.~Huang, Z.~Chen, S.~Kowalski, Y.~G. Ma, R.~Wada, T.~Keutgen, K.~Hagel,
  M.~Barbui, A.~Bonasera, and C.~Bottosso {\em et al.,}.
\newblock {\em Phys. Rev. C}, 81:044620, 2010.

\bibitem{Su11}
{Jun Su}, {Feng-Shou Zhang}, and {Bao-An Bian}.
\newblock {\em Phys. Rev. C}, 83:014608, 2011.

\bibitem{Mei14}
B.~Mei, H.~S. Xu, X.~L. Tu, Y.~H. Zhang, Yu.~A. Litvinov, K.-H. Schmidt,
  M.~Wang, Z.~Y. Sun, X.~H. Zhou, Y.~J. Yuan, M.~V. Ricciardi, and
  A.~Kelic-Heil {\em et al.,}.
\newblock {\em Phys. Rev. C}, 89:054612, 2014.

\bibitem{Jensen84}
A.~S. Jensen, P.~G. Hansen, and B.~Jonson.
\newblock {\em Nucl. Phys. A}, 431:393, 1984.

\bibitem{Hove13}
D.~Hove, A.~S. Jensen, and K.~Riisager.
\newblock {\em Phys. Rev. C}, 88:064329, 2013.

\bibitem{Iancu05}
G.~Iancu, F.~Flesch, and W.~Heinrich.
\newblock {\em Radiat. Meas.}, 39:525, 2005.

\bibitem{Cheng12}
J.~X. Cheng, X.~Jiang, S.~W. Yan, and D.~H. Zhang.
\newblock {\em J. Phys. G}, 39:055104, 2012.

\bibitem{minuit}
F.~James.
\newblock "\textsc{MINUIT}: {F}unction {M}inimization and {E}rror {A}nalysis -
  {R}eference {M}anual".
\newblock http://wwwasdoc.web. cern.ch/wwwasdoc/minuit/minmain.html, 2000.

\bibitem{ranlux}
M.~Luescher.
\newblock "{A} portable high-quality random number generator for lattice field
  theory calculations".
\newblock {\em Computer Physics Communications}, 79, 1994.

\bibitem{Boole1860}
G.Boole.
\newblock {\em A treatise on the calculus of finite differences}.
\newblock Cambridge University Press, New York, 2009.
\newblock Reproduction of the original edition published in 1860.

\bibitem{Jordan1950}
C.Jordan.
\newblock {\em Calculus of finite differences}.
\newblock Schaum's Outlines. Chelsea Publishing Comp., New York, 1950.

\bibitem{Spiegel1970}
M.R.Spiegel.
\newblock {\em Calculus of finite differences and difference equations}.
\newblock McGraw-Hill, Inc., 1971.

\end{thebibliography}

\end{document}